\documentclass[a4paper,10pt]{article}\textheight 24.5cm \textwidth 17cm\voffset=-1.in\hoffset=
-1.in

\makeatletter \@addtoreset{equation}{section} \makeatother

\usepackage[dvips]{graphicx}
\usepackage{amsmath}
\usepackage{graphicx}
\usepackage{amsfonts}
\usepackage{amscd}
\usepackage{amssymb}
\usepackage{psfrag}
\usepackage{hhline}
\usepackage{cite}
\usepackage{upgreek}
\usepackage{epsfig,amsfonts}

\usepackage{mathtext}         %

\begin{document}

%\draft
%%%%%%%%%%%%%%%%%%% Title %%%%%%%%%%%%%%%%%%%%%%%%%%%%%%%%%%%%%%%%%%%

\vspace{0.5cm}

\begin{center}
\begin{LARGE}
{\Large \textbf{Five-point Correlation Numbers in One-Matrix Model}}

\vspace{0.3cm}

\end{LARGE}

\vspace{0.5cm}

\vspace{1.3cm}

\begin{large}

\textbf{G.Tarnopolsky\footnote{hetzif@itp.ac.ru}}

\end{large}
\large

 \vspace{1.cm}

L.D. Landau Institute for Theoretical Physics\\
  Chernogolovka, 142432, Russia\\

\vspace{1.0cm}

\centerline{\bf Abstract}

\vspace{.8cm}

\parbox{11cm}
{The five-point correlation numbers in the One-matrix model is
calculated in the Liouville frame. Validity of the fusion rules for
it is checked.}
\end{center}

\bigskip

\large

\section{Introduction}

There exist two approaches to the 2d Quantum gravity. One of them is
the continuous approach. In this approach the theory is determined
by the functional integral over all metrics \cite{polyakov}.
Calculation of this integral in conformal gauge leads to the
Liouville field theory. Therefore this approach is called the
Liouville gravity.

The other way to describe sum over all surfaces is  the discrete
approach. It is based on the idea of approximation of
two-dimensional geometry by the ensemble of planar graphs of big
size. Technically the ensemble of graphs is usually defined by
expansion into a series of perturbation theory of integral over
matrixes. That is why this approach is called the Matrix models.
References to the both approaches can be found in the review
\cite{review}.

The main objects in these two approaches are the correlation numbers
$\langle O^{L}_{1}...O^{L}_{N} \rangle$ of the observables
$O^{L}_{k}$ in the Liouville gravity and the correlation numbers
$\langle O^{M}_{1}...O^{M}_{N} \rangle$ of the observables
$O^{M}_{k}$ in the Matrix models. It was a remarkable  discovery
that the spectra of the gravitational dimensions \cite{review} in
the both approaches coincide. It was reason to assume that both
these theories describe the same variant of the 2d Quantum gravity
and therefore the correlation numbers will coincide. However  the
attempt of naive identification of the correlation numbers  doesn't
lead to the coincidence in general case.

It was remarked in \cite{MSS}, \cite{BZ2}, that existence of the
so-called resonances makes the identification of the correlation
numbers ambiguous. In order to investigate this ambiguity it is
convenient to pass from the correlation numbers to generating
functions of correlators according to the formulas
$$
F^{L}(\lambda_{1},...,\lambda_{N}) = \langle \exp(\sum
\lambda_{k}O_{k}^{L}) \rangle , \quad \mathcal{Z}(t_{1},...,t_{N}) =
\langle \exp (\sum t_{k}O_{k}^{M}) \rangle.
$$
In \cite{MSS}, \cite{BZ2} a conjecture was proposed  that there
exists a ``resonance'' transformation $t_{k}=t_{k}(\{\lambda\})$
(from KdV frame to Liouville frame), such that the correlation
functions in the both theories will coincide
$$
F^{L}(\lambda_{1},...,\lambda_{N}) =
\mathcal{Z}(t_{1}(\{\lambda\}),...,t_{N}(\{\lambda\})).
$$
The form of the transformation was conjectured in \cite{BZ2} for
particular case, namely for the Minimal quantum gravity
$\mathcal{MG}_{2/2p+1}$ and the One-matrix model with p critical
points. In loc. cit. the conjectured of the coincidence was checked
up to four-point correlator.

In this paper we continue comparing of $\mathcal{MG}_{2/2p+1}$  and
the One-matrix model. Using the ``resonance'' \; transformation
\cite{BZ2} we find the five-point correlation numbers in One-matrix
model in the Liouville frame. And we checked that the correlation
numbers satisfy the fusion rules, which necessarily must be
satisfied in the Minimal gravity $\mathcal{MG}_{2/2p+1}$.

The article is organized in the following way. In the first part of
the article there is a brief summary of the results of  the  paper
\cite{BZ2}. In the second part, an expression for  the five-point
correlation numbers in the Liouville frame is found, and validity of
the fusion rules for it is proved.

\section{One-matrix model}
As it has been shown in  the classical works on  the Matrix models
(see review \cite{review}) in the scaling limit near the
$p$-critical point the partition function of the One-matrix model
can be described in terms of the solution of the ``string equation''
\begin{eqnarray}
\mathcal{P}(u)=0, \label{Str0}
\end{eqnarray}
where $\mathcal{P}(u)$ is the polynomial of degree $p+1$ ($p$ is
natural number)
\begin{eqnarray}
\mathcal{P}(u)=u^{p+1}+t_{0}u^{p-1}+\sum\limits_{k=1}^{p-1}t_{k}u^{p-k-1},
\label{Pol}
\end{eqnarray}
with the parameters $t_{k}$ describing deviation from the
$p$-critical point. The singular part of the partition function in
the Matrix models $\mathcal{Z}(t_{0}, t_{1},...t_{p-1})$ can be
described according to (\ref{Pol}), as
\begin{eqnarray}
\mathcal{Z} = \frac{1}{2}\int_{0}^{u_{*}}\mathcal{P}^{2}(u)du,
\label{StS}
\end{eqnarray}
where $u_{*}=u_{*}(t_{0},t_{1},...,t_{p-1})$ is the maximal real
root \cite{BZ2}  of the  polynomial (\ref{Pol}). Expression
(\ref{StS}) gives only the singular part of the partition function
in the Matrix models, but complete the matrix integral includes also
the regular part, which is analytical with respect to all parameters
$t_{k}$ at the point $\{t_{1},...,t_{p-1}\}=0$. It is not worth
considering it. If we calculate the correlation numbers by taking
derivatives in the parameters $t_{k}$ (KdV frame), according to the
formula
\begin{eqnarray}
\langle O^{M}_{k_{1}}O^{M}_{k_{2}}...O^{M}_{k_{N}}
\rangle=\mathcal{Z}_{k_{1}...k_{N}}=\left.\frac{\partial^{N}\mathcal{Z}}{\partial
t_{k_{1}}...\partial t_{k_{N}}}\right|_{t_{1}=...=t_{p-1}=0},
\label{Corls}
\end{eqnarray}
that coincidence with results of the Minimal gravity
$\mathcal{MG}_{2/2p+1}$ will not take place.

In the work \cite{BZ2} it was argued that among analytic and
scale-invariant transformations $t_{k}=t_{k}(\{\lambda\})$, exists
the special one, after which the  correlation numbers satisfy the
fusion rules, which necessarily must be satisfied in the Minimal
gravity $\mathcal{MG}_{2/2p+1}$ . The fusion rules are expressed as
follows
\begin{equation}
\langle O_{k_{1}}O_{k_{2}}...O_{k_{N}} \rangle=0, \quad
\textrm{if}\quad
\begin{cases}
k_{1}+...+k_{N-1}<k_{N}, \;\;\quad \textrm{when} \quad k_{1}+...+k_{N} \quad \textrm{is even}, \\
k_{1}+...+k_{N}<2p-1 , \quad \textrm{when} \quad k_{1}+...+k_{N}
\quad \textrm{is odd},
\end{cases} \label{fusion}
\end{equation}
Here it is assumed that  $k_{i}$ run through  the range
$k_{i}=0,1,...,p-1$ and that $k_{N}$ is the maximal index, i.e.
$k_{i} \leqslant k_{N}$. Later we will say that we are in even(odd)
sector, if $k_{1}+...+k_{N}$ is even(odd). After this transformation
the polynomial $\mathcal{P}(u)$, up to  the  factor
$\frac{(p+1)!}{(2p-1)!!}u_{0}^{p+1}$ takes the form \cite{BZ2}:
\begin{eqnarray}
Q(x,\{s\})
=\sum\limits_{n=0}^{\infty}\sum\limits_{k_{1}...k_{n}=1}^{p-1}
\frac{s_{k_{1}}...s_{k_{n}}}{n!}\frac{d^{n-1}} {dx^{n-1}}L_{p-\sum
k_{i}-n}(x),
\end{eqnarray}
here, we already pass from dimensional Liouville parameters
$\{\lambda_{k}\}$ to dimentionless $\{s_{k}\}$, by the formula
\begin{eqnarray}
s_{k} = \frac{g_{k}}{g}\frac{u_{0}^{-k-2}}{2p+1}\lambda_{k}, \quad
\textrm{where} \quad g_{k}= \frac{(p-k-1)!}{(2p-2k-3)!!}, \quad
g=\frac{(p+1)!}{(2p+1)!!}.
\end{eqnarray}
Also $x=u/u_{0}$, where $u_{0}$ is $u_{*}$ at $s_{1},...,s_{p-1}=0$.
The relation between $t_{0}$ and $u_{0}$ is
\begin{eqnarray}
t_{0}= -\frac{1}{2}\frac{p(p+1)}{2p-1}u_{0}^{2}.
\end{eqnarray}
And $L_{n}(x)$ are the Legendre polynomials (see Appendix A). We
 assume that $\left(\frac{d}{dx}\right)^{-1}L_{p} = \int L_{p}
dx =\frac{L_{p+1}-L_{p-1}}{2p+1}$. Below we use the notations
\begin{eqnarray}
Q_{k_{1}...k_{n}}(x)=\frac{d^{n-1}} {dx^{n-1}}L_{p-\sum k_{i}-n}(x),
\quad Q_{0}(x)= \frac{L_{p+1}-L_{p-1}}{2p+1}.
\end{eqnarray}
This ``resonance''  transformation is expressed as
\begin{align}
t_{k}&=
\frac{g}{g_{k}}\frac{2p+1}{u_{0}^{-k-2}}\cdot\sum_{n=1}^{[\frac{2+k}{2}]}
\sum_{\substack{
   m_{1},...,m_{n}=0 \\m_{1}+...+m_{n}=\\=k+2-2n}
   }^{p-1}\frac{(2p-2k-5+2n)!!}{(2p-2k-3)!!}\cdot
   \frac{s_{m_{1}}...s_{m_{n}}}{n!},
\end{align}
where we introduce auxiliary parameter $s_{0}=-\frac{1}{2}$.

Thus  the  singular part of the partition function in  the Matrix
models expressed in terms of the new parameters can be written as
\begin{eqnarray}
\mathcal{Z} = \frac{1}{2}\int_{0}^{x_{*}}Q^{2}(x)dx, \label{ZQ}
\end{eqnarray}
where $x_{*}=x_{*}(s_{1},...,s_{p-1})$ is the maximal real root of
the  polynomial $Q(x)$, and $x_{*}(0,0,...,0)=1$. The correlation
numbers in the Liouville frame  are defined as follows
\begin{eqnarray}
\langle O^{M}_{k_{1}}O^{M}_{k_{2}}...O^{M}_{k_{N}}
\rangle=\mathcal{Z}_{k_{1}...k_{N}}=\left.\frac{\partial^{N}\mathcal{Z}}{\partial
s_{k_{1}}...\partial s_{k_{N}}}\right|_{s_{1}=...=s_{p-1}=0}.
\label{Corls}
\end{eqnarray}
These correlators  will coincide with those of  the  Minimal gravity
$\mathcal{MG}_{2/2p+1}$. In  the next section we will give  the
answers derived from the formula (\ref{Corls}) for   the three- and
four-point correlation numbers.

\subsection{Three- and four-point correlation numbers}

In the next sections we will use convenient notations
\begin{eqnarray}
k = \sum\limits_{i=1}^{N}k_{i}, \quad \textrm{and}\quad
k_{i_{1}...i_{n}}^{j_{1}...j_{m}} =
(k_{i_{1}}+...+k_{i_{n}})-(k_{j_{1}}+...+k_{j_{m}}). \label{Oboz2}
\end{eqnarray}
To simplify  expressions we need a symbol of symmetrization, denoted
by parentheses, for example
\begin{eqnarray}
Q_{(k_{1}k_{2}}Q_{k_{3})}
=Q_{k_{1}k_{2}}Q_{k_{3}}+Q_{k_{1}k_{3}}Q_{k_{2}}+Q_{k_{2}k_{3}}Q_{k_{1}},
\end{eqnarray}
notice, that  the order of indexes in each $Q$-term doesn't matter.

The answer for  the three-point correlation numbers can be written
in  the form
\begin{multline}
\mathcal{Z}_{k_{1}k_{2}k_{3}} = \frac{1}{2}\int_{-1}^{1}
(Q_{(k_{1}k_{2}}Q_{{k_{3}})}+ Q_{0}Q_{k_{1}k_{2}k_{3}})dx-\\
-\frac{Q_{k_{1}}(1)Q_{k_{2}}(1)Q_{k_{3}}(1)}{Q'_{0}(1)}-\frac{Q_{0}(1)
Q_{k_{1}k_{2}}(1)Q_{k_{3}}(1)}{Q'_{0}(1)}. \label{Zkkk}
\end{multline}
Taking into account  the values of the Legendre polynomials (see
Appendix A) and their derivatives in the  point $x=1$, and also that
$Q_{0}(x)$ is orthogonal to $Q_{k_{1}k_{2}k_{3}}(x)$, we get
\begin{eqnarray}
\mathcal{Z}_{k_{1}k_{2}k_{3}} = -1+\frac{1}{2}\int_{-1}^{1}
Q_{(k_{1}k_{2}}Q_{{k_{3}})} dx. \label{Zzkkk}
\end{eqnarray}
Now we give  the  general final answer for
$\mathcal{Z}_{k_{1}k_{2}k_{3}}$, assuming that $0 \leqslant k_{1}
\leqslant k_{2} \leqslant k_{3} \leqslant p-1$
\begin{eqnarray}
\mathcal{Z}_{k_{1}k_{2}k_{3}} =
\begin{cases}
-1, \quad \textrm{if} \quad  k_{3}\leqslant k_{12},\\
\;\;\, 0, \quad \textrm{if} \quad   k_{3}> k_{12},
\end{cases}
\end{eqnarray}
in the even sector, and
\begin{eqnarray}
\mathcal{Z}_{k_{1}k_{2}k_{3}} =
\begin{cases}
-1, \quad \textrm{if} \quad k \geqslant 2p-1, \\
\textrm{reg.}, \;\,\; \textrm{if} \quad k < 2p-1,
\end{cases}
\end{eqnarray}
in the odd sector, where $"\textrm{reg.}"$ are  the regular terms.

Here we give the answer for four-point correlation numbers
\begin{multline}
\mathcal{Z}_{k_{1}k_{2}k_{3}k_{4}} = \frac{1}{2}\int_{-1}^{1}
\left(Q_{(k_{1}} Q_{k_{2}k_{3}k_{4})}+Q_{(k_{1}k_{2}}
Q_{k_{3}k_{4})}+Q_{0}Q_{k_{1}k_{2}k_{3}k_{4}}
\right)dx- \\
-\frac{Q_{(k_{1}k_{2}}Q_{k_{3}}Q_{k_{4})}}{Q'_{0}}+
\frac{Q'_{(k_{1}}Q_{k_{2}}Q_{k_{3}}Q_{k_{4})}}{(Q'_{0})^{2}}-
\frac{Q''_{0}Q_{k_{1}}Q_{k_{2}}Q_{k_{3}} Q_{k_{4}}}{(Q'_{0})^{3}},
\label{Zkkkk1}
\end{multline}
where all $Q$-terms after  the integral are taken in the  point
$x=1$. Now assuming, as usual, that $ 0 \leqslant k_{1} \leqslant
k_{2} \leqslant k_{3} \leqslant k_{4} \leqslant p-1, $ in  the even
sector we have
\begin{eqnarray}
\mathcal{Z}_{k_{1}k_{2}k_{3}k_{4}} =
\begin{cases}
\frac{1}{2}(3p^{2}-5p-(2p-1)k +\sum k_{i}^{2} ), \quad\qquad
\qquad\;  p-1 < k_{12}\\
(1+k_{1})(2p-3-k)+F(k_{14})+F(k_{13}), \qquad k_{12}\leqslant p-1<k_{13},\\
(1+k_{1})(2p-3-k)+F(k_{14}),\qquad \qquad \qquad \,  k_{13}\leqslant p-1<k_{14},k_{23},\\
(1+k_{1})(2p-3-k), \qquad\qquad \qquad \qquad\qquad\; k_{14}<
k_{23}, \; k_{14} \leqslant p-1,\\
\frac{1}{2}(2+k_{123}-k_{4})(2p-3-k) ,\qquad\qquad\qquad\;\;\;\,  k_{14}\geqslant k_{23}, \; k_{23} \leqslant p-1,\\
0,\qquad\qquad\qquad\qquad \qquad\qquad \qquad\qquad\qquad\;\; k_{4}-k_{123}> 0,\\
\end{cases}
\end{eqnarray}
where $F(k)=\frac{1}{2}(p-k-1)(p-k-2)$. And in the odd sector
\begin{eqnarray}
\mathcal{Z}_{k_{1}k_{2}k_{3}k_{4}} =
\begin{cases}
\frac{1}{2}(3p^{2}-5p-(2p-1)k +\sum k_{i}^{2} ), \qquad\qquad\quad\; p-1 < k_{12},\\
(1+k_{1})(2p-3-k)+F(k_{14})+F(k_{13}), \qquad k_{12}\leqslant p-1<k_{13}, \\
(1+k_{1})(2p-3-k)+F(k_{14}),\qquad\qquad\qquad\,  k_{13}\leqslant p-1<k_{14},k_{23},\\
(1+k_{1})(2p-3-k), \qquad\qquad\qquad\qquad \qquad\;   k_{14}< k_{23},\; k_{14} \leqslant p-1, \; k \geqslant 2p-3, \\
\frac{1}{2}(2+k_{123}-k_{4})(2p-3-k), \qquad\qquad\qquad\;\;\;
k_{14}\geqslant k_{23},
\; k_{23} \leqslant p-1,\; k \geqslant 2p-3,\\
\textrm{reg. }.  \qquad\qquad\qquad\qquad \qquad\qquad\qquad\qquad \quad k \leqslant 2p-5.\\
\end{cases}
\end{eqnarray}
One can see, that near  the "critical" regions, i.e.  the regions in
which relations between $k_{i}$ are already similar to those for
which  the fusion rules must be valid, the  correlation numbers are
factorized and become  the simple form. In  other regions  the
correlators has very bulky form. This interesting feature of  the
"simplification" of  the correlator will be noted in  the
calculation of the  five-point correlation numbers.

\section{Five-point correlation numbers }
We can at last begin to investigate  the five-point correlation
numbers in  the  Matrix models. The answer for it can be written as
follows
\begin{align}
\mathcal{Z}_{k_{1}k_{2}k_{3}k_{4}k_{5}}  =& \frac{1}{2}\int_{-1}^{1}
\left(Q_{(k_{1}} Q_{k_{2}k_{3}k_{4}k_{5})}+
Q_{(k_{1}k_{2}}Q_{k_{3}k_{4}k_{5})}
+Q_{0}Q_{k_{1}k_{2}k_{3}k_{4}k_{5}}\right)dx-
\frac{Q_{(k_{1}k_{2}k_{3}}Q_{k_{4}}Q_{k_{5})}}
{Q_{0}'}-\notag\\
&-\frac{Q_{(k_{1}k_{2}}Q_{k_{3}k_{4}} Q_{k_{5})}}{Q_{0}'}+
\frac{Q'_{(k_{1}k_{2}}Q_{k_{3}}Q_{k_{4}} Q_{k_{5})}}{(Q_{0}')^{2}}
+\frac{Q'_{(k_{1}}Q_{k_{2}k_{3}} Q_{k_{4}}Q_{k_{5})}}{(Q'_{0})^{2}}-
\notag\\
&-\frac{Q''_{(k_{1}}Q_{k_{2}}Q_{k_{3}} Q_{k_{4}}Q_{k_{5})}}
{(Q'_{0})^{2}}- \frac{2Q'_{(k_{1}}Q'_{k_{2}}Q_{k_{3}}
Q_{k_{4}}Q_{k_{5})}} {(Q'_{0})^{3}}-
Q''_{0}\frac{Q_{(k_{1}k_{2}}Q_{k_{3}}Q_{k_{4}}
Q_{k_{5})}}{(Q'_{0})^{3}}+\notag\\
&+3Q''_{0} \frac{Q'_{(k_{1}}Q_{k_{2}}Q_{k_{3}}Q_{k_{4}}
Q_{k_{5})}}{(Q'_{0})^{4}}+\left(Q'''_{0}-\frac{3(Q''_{0})^{2}}{Q'_{0}}\right)
\frac{Q_{k_{1}}Q_{k_{2}} Q_{k_{3}}Q_{k_{4}}Q_{k_{5}}}{(Q'_{0})^{4}},
\label{Zkkkkk1}
\end{align}
where all   $Q$-terms after  the integral are taken in  the  point
$x=1$. After simplification (see Appendix C), one can get
\begin{align}
\mathcal{Z}_{k_{1}k_{2}k_{3}k_{4}k_{5}} =& \frac{1}{2}\int_{-1}^{1}
\left(Q_{(k_{1}}Q_{k_{2}k_{3}k_{4}k_{5})}+
Q_{(k_{1}k_{2}}Q_{k_{3}k_{4}k_{5})}\right)dx+
\left(Q'''_{0}-3(Q''_{0})^{2}\right)-\notag\\
&+\sum\limits_{i=1}^{5}(3Q''_{0}Q'_{k_{i}}-Q''_{k_{i}})+
\sum\limits_{i<j}(Q'_{k_{i}k_{j}}-
2Q'_{k_{i}}Q'_{k_{j}}-Q''_{0}Q_{k_{i}k_{j}})+
 \notag\\
&+\sum\limits_{i,j,l}Q'_{k_{i}}Q_{k_{j}k_{l}}-\sum\limits_{i<j<l}Q_{k_{i}k_{j}k_{l}}-
\sum\limits_{i,j,k,l}Q_{k_{i}k_{j}}Q_{k_{l}k_{m}}.
\end{align}
For convenience let us divide  the  five-point correlator in several
parts
\begin{eqnarray}
\mathcal{Z}_{k_{1}k_{2}k_{3}k_{4}k_{5}}=
\mathcal{Z}_{k_{1}k_{2}k_{3}k_{4}k_{5}}^{(\textrm{I})}+
\mathcal{Z}_{k_{1}k_{2}k_{3}k_{4}k_{5}}^{(\textrm{J})}+
\mathcal{Z}_{k_{1}k_{2}k_{3}k_{4}k_{5}}^{(\textrm{0})},
\end{eqnarray}
where  the first two integral terms are
\begin{eqnarray}
\mathcal{Z}_{k_{1}k_{2}k_{3}k_{4}k_{5}}^{(\textrm{I})}
=\frac{1}{2}\int_{-1}^{1} Q_{(k_{1}}Q_{k_{2}k_{3}k_{4}k_{5})}dx,
\qquad  \mathcal{Z}_{k_{1}k_{2}k_{3}k_{4}k_{5}}^{(\textrm{J})}
=\frac{1}{2}\int_{-1}^{1} Q_{(k_{1}k_{2}}Q_{k_{3}k_{4}k_{5})}dx,
\end{eqnarray}
and the last term is given by the formula
\begin{align}
\mathcal{Z}_{k_{1}k_{2}k_{3}k_{4}k_{5}}^{(\textrm{0})} =
&\left(Q'''_{0}-3(Q''_{0})^{2}\right)+
\sum\limits_{i=1}^{5}(3Q''_{0}Q'_{k_{i}}-Q''_{k_{i}})
-\sum\limits_{i<j} 2Q'_{k_{i}}Q'_{k_{j}}+ \notag\\
&+\sum\limits_{i<j}(Q'_{k_{i}k_{j}}-Q''_{0}Q_{k_{i}k_{j}}) -
\sum\limits_{i<j<l}Q_{k_{i}k_{j}k_{l}}+
\sum\limits_{i,j,l}Q'_{k_{i}}Q_{k_{j}k_{l}}-
\sum\limits_{i,j,k,l}Q_{k_{i}k_{j}}Q_{k_{l}k_{m}}
\end{align}
After  the  further simplification  (see Appendix C), one can find
\begin{multline}
\mathcal{Z}_{k_{1}k_{2}k_{3}k_{4}k_{5}}^{(\textrm{0})} =
\\
=\frac{1}{8}(4\sum_{i=1}^{5}k_{i}^{2}-k^{2}-2k -8-
\sum\limits_{m<n}k_{ijl}^{mn}(k_{ijl}^{mn}+2)
\Theta(k_{mn}-p))(2p-3-k)(2p-5-k)+\\
+\sum\limits_{i<j<l}H(k_{ijl})G_{1}-\sum\limits_{i,j,l,m}F(k_{ij})F(k_{lm})G_{2},
\label{Corup}
\end{multline}
where $H(k)=\frac{1}{8}\prod\limits_{r=1}^{4}(p-r-k)$, and also two
factors $G_{1}=\Theta(k_{ijl}-p)+ \Theta(k_{mn}-p)-1 \;$ and
$G_{2}=\Theta(k_{ij}-p)\Theta(k_{lm}-p)$, and $\Theta(a-b)$ is the
step-function, which is defined as
\begin{eqnarray}
\Theta(a-b)=
\begin{cases}
1, \quad \textrm{if}\quad a>b, \\
0, \quad \textrm{if}\quad  a\leqslant b.
\end{cases}
\end{eqnarray}

Now we get down to consideration of  the  correlation numbers in the
particular sectors. Let us begin with  the odd sector.

\subsection{Odd Sector}
In this sector $k=k_{1}+k_{2}+k_{3}+k_{4}+k_{5}$ is odd, and $
\mathcal{Z}_{k_{1}k_{2}k_{3}k_{4}k_{5}}^{(\textrm{I})} =
\mathcal{Z}_{k_{1}k_{2}k_{3}k_{4}k_{5}}^{(\textrm{J})}=0 $ due to
oddness of   the integrand. Thus one can get
\begin{multline}
\mathcal{Z}_{k_{1}k_{2}k_{3}k_{4}k_{5}}^{(\textrm{odd})} =
\mathcal{Z}_{k_{1}k_{2}k_{3}k_{4}k_{5}}^{(\textrm{0})}=\\
=\frac{1}{8}(4\sum_{i=1}^{5}k_{i}^{2}-k^{2}-2k -8-
\sum\limits_{m<n}k_{ijl}^{mn}
(k_{ijl}^{mn}+2)\Theta(k_{mn}-p))(2p-3-k)(2p-5-k)+\\
+\sum\limits_{i<j<l}H(k_{ijl})G_{1}-\sum\limits_{i,j,l,m}F(k_{ij})F(k_{lm})G_{2}.
\label{zodd}
\end{multline}
If $k \leqslant 2p -7$,  the  terms of  the  partition function are
regular \cite{BZ2}. Thus, due to the formula (\ref{fusion}) we only
need to check  the fusion rules at  the points $k=2p-5$ and
$k=2p-3$. It is obvious, that if $k=2p-3$, and $k=2p-5$, the first
term in the r.h.s. of the formula (\ref{zodd}) equals zero. Also
notice that $G_{2}=0$, because $G_{2} = 1$ at least if $k\geqslant
2p$. For $k=2p-3, 2p-5$, $G_{1}=0$ or $G_{1}=-1$. In case $G_{1}=-1$
we have the inequalities $k_{mn}< p$ and $k_{ijl}< p$, therefore
$k_{ijl}$ can be equal only $p-4,\, p-3, \, p-2, \, p-1$, but for
these values $H(k_{ijl}) =0$. Thus we showed, that the second term
in the formula (\ref{zodd}) also equals zero for $k=2p-3$ and
$k=2p-5$.

As  the  result we proved  the validity of the  fusion rules for
$\mathcal{Z}_{k_{1}k_{2}k_{3}k_{4}k_{5}}^{(\textrm{odd})}$, i.e.
\begin{eqnarray}
\mathcal{Z}_{k_{1}k_{2}k_{3}k_{4}k_{5}}^{(\textrm{odd})}=0, \quad
\textrm{when} \quad k<2p-1.
\end{eqnarray}
Now let us pass to consideration of  the even sector.

\subsection{Even Sector}
In  the even sector  $k=k_{1}+k_{2}+k_{3}+k_{4}+k_{5}$ is even.
After calculation
$\mathcal{Z}_{k_{1}k_{2}k_{3}k_{4}k_{5}}^{(\textrm{J})}$ (see
formula (\ref{I2})) and summation of it with
$\mathcal{Z}_{k_{1}k_{2}k_{3}k_{4}k_{5}}^{(0)}$ one can get
\begin{multline}
\mathcal{Z}_{k_{1}k_{2}k_{3}k_{4}k_{5}}^{(\textrm{even})} =
\mathcal{Z}_{k_{1}k_{2}k_{3}k_{4}k_{5}}^{(\textrm{I})}+\\
+\frac{1}{8}(4\sum_{i=1}^{5}k_{i}^{2}-k^{2}-2k -8-
\sum\limits_{m<n}G_{3}k_{ijl}^{mn}(k_{ijl}^{mn}+2))(2p-3-k)(2p-5-k)+\\+
\left(\sum\limits_{i<j<l}H(k_{ijl})-\sum\limits_{i,j,l,m}F(k_{ij})F(k_{lm})
\right)G_{2},
\end{multline}
where $G_{3}=\Theta(k_{mn}-p)+\Theta(k_{mn}-k_{ijl}-2)
\Theta(p-1-k_{ijl})\Theta(p-1-k_{mn})$ and
$G_{2}=\Theta(k_{ij}-p)\Theta(k_{lm}-p)$.

We as usual assume  the following ordering $ 0 \leqslant k_{1}
\leqslant k_{2} \leqslant k_{3}\leqslant k_{4}\leqslant k_{5}
\leqslant p-1 \label{Usl5}. $ If $k_{5} > k_{1234}$ it is obvious,
that $G_{2}=0$. Then by means of  the simple reasoning one can
prove, that $G_{3}=0$, when one of the indexes $i,j,l$ equals five.
In remaining cases $G_{3}=1$. Having calculated
$\mathcal{Z}_{k_{1}k_{2}k_{3}k_{4}k_{5}}^{(\textrm{I})}$ (see
formula (\ref{I1})), we derive  the following formula
\begin{eqnarray}
\mathcal{Z}_{k_{1}k_{2}k_{3}k_{4}k_{5}}^{(\textrm{even})} =
\frac{1}{8}(k_{5}^{1234}-2)(k_{5}^{1234}-4)(2p-3-k)(2p-5-k)
(1-\Theta(k_{5}^{1234}-6)) \label{zeven}.
\end{eqnarray}
From this formula we can see
$\mathcal{Z}_{k_{1}k_{2}k_{3}k_{4}k_{5}}^{(\textrm{even})}=0$ if
$k_{5}> k_{1234}$. Thus the  fusion rules are valid. Notice that
when $k_{5}=k_{1234}+2$ and $k_{5}=k_{1234}+4$ nulling of  the
function take place automatically, without  the integral term
$\mathcal{Z}_{k_{1}k_{2}k_{3}k_{4}k_{5}}^{(\textrm{I})}$.

\section{Conclusion}
In this paper  the  five-point correlation numbers have been
calculated. These numbers are necessary for
the several reasons. First it is one more test of  the fusion rules.

Second, as we assume  the correlation
numbers in  the two approaches are the same.
Therefore the five-point correlation numbers which were found in this paper have
to coincide with answers in continious approach. But the five-point correlation numbers in
continious approach are still unknown.

\section{Aknowledgements}

I am grateful to A.A. Belavin for posing the problem and attention
to this work. I am also grateful to Y. Pugai, M.Lashkevich and
M.Bershtein for useful discussion. This work was supported by
RBRF-CNRS grant PICS-09-02-91064 and by RFBR initiative
interdisciplinary project grant 09-02-12446-OFI-m and by the Russian
Ministry of Science and Technology under Scientific Schools grant
3472.2008.2. The research was held within the bounds of Federal
Program "Scientific and Scinetific-Pedagogical personnel of
innovational Russia" on 2009-2013 , goskontrakt N P1339.

\section*{Appendix}
\appendix

\section{Legendre Polynomials}

The Legendre polynomials $L_{n}(x)$ are $n$-th order polynomials,
which form an orthogonal system on the interval $[-1,1]$ with the
weight $1$,
\begin{eqnarray}
\int_{-1}^{1}L_{n}(x)L_{m}(x) dx =\frac{2\delta_{m,n}}{2n+1}.
\end{eqnarray}
The standard normalization is such that
\begin{eqnarray}
L_{n}(1)=1.
\end{eqnarray}
Explicit formula for $L_{n}(x)$ is
\begin{eqnarray}
L_{n}(x) = \frac{2^{-n}}{n!}\frac{d^{n}}{dx^{n}}[x^{2}-1]^{n} =
2^{-n}\sum\limits_{l=0}^{[n/2]}(-1)^{l}\frac{(2n-2l)!}{l!(n-l)!(n-2l)!}x^{n-2l}.
\end{eqnarray}
Expression of polynomial $L_{n}(x)$ in terms of Hypergeometric is
series
\begin{eqnarray}
L_{n}(x)={}_{2}F_{1}\left(-n,n+1,1;\frac{1-x}{2}\right),
\end{eqnarray}
from this we can obtain
\begin{eqnarray}
L'_{n}(1) = \frac{n(n+1)}{2}, \quad L''_{n}(1) =
\frac{(n-1)n(n+1)(n+2)}{8}, \quad \textrm{etc}
\end{eqnarray}
Yet another closed form is in terms of the contour integral
\begin{eqnarray}
L_{n}(x) = \oint_{0} \frac{(1-2xz
+z^{2})^{-1/2}}{z^{n+1}}\frac{dz}{2\pi i}.
\end{eqnarray}

The following relations are useful in our analysis:
\begin{eqnarray}
L'_{n+1}(x)-L'_{n-1}(x) = (2n+1)L_{n}(x),
\end{eqnarray}
they are valid for all $n=0,1,2,3...$if one assume, that
$L_{-1}(x)=0$. The following formulas  are
\begin{eqnarray}
\frac{1}{2}\int_{-1}^{1}L'_{n}(x)L_{m}(x)dx =
E_{n+m-1}\Theta_{n,m+1},
\end{eqnarray}
\begin{eqnarray}
\frac{1}{2}\int_{-1}^{1}L''_{n}(x)L_{m}(x)dx =
E_{n+m}\Theta_{n,m+2}\frac{(n+m+1)(n-m)}{2}, \label{L''L}
\end{eqnarray}
and in general
\begin{eqnarray}
\frac{1}{2}\int_{-1}^{1}L^{(l)}_{n}(x)L_{m}(x)dx=
E_{n+m-l}\Theta_{n,m+l}\frac{2^{-l+1}}{(l-1)!}\prod\limits_{s=0}^{l-2}(n+m+l-1-2s)
(n-m+l-2-2s), \label{LlL}
\end{eqnarray}
where $L^{(l)}_{n}(x)$ stands for the $l$-th derivative. Here
$\Theta_{n,m}=L_{n-m}(1)$ is the step function, and
\begin{eqnarray}
E_{n}=
\begin{cases}
1,\quad \textrm{if}\quad n \textrm{\;\;is even},  \\
0, \quad \textrm{if}\quad n \textrm{\;\;is odd}.
\end{cases}
\end{eqnarray}
Integrating (\ref{L''L}) by parts, we have
\begin{eqnarray}
\frac{1}{2}\int_{-1}^{1}L'_{n}(x)L'_{m}(x)dx =
E_{n+m}\left[\Theta_{m,n}\frac{n(n+1)}{2}+\Theta_{n,m}\frac{m(m+1)}{2}\right].
\end{eqnarray}
The general formula which expresses   $m$-th derivative of  the
Legendre polynomial of $n$-th order, by  the  sum of  the Legendre
polynomials is
\begin{eqnarray}
\frac{d^{m}}{dx^m}L_{n}(x)=\sum\limits_{k:2}^{n-m}(2k+1)B^{(m)}
_{n,k}L_{k}(x), \label{Main}
\end{eqnarray}
where
\begin{eqnarray}
 B^{(m)} _{n,k}=
\frac{2^{m-1}}{(m-1)!}\frac{\Gamma \left(\frac{n+k+m+1}{2}\right)
\Gamma \left(\frac{n-k+m}{2}\right)}{\Gamma
\left(\frac{n+k-m+3}{2}\right) \Gamma \left(\frac{n-k-m+2}{2}\right)
}.
\end{eqnarray}
notice, that (\ref{LlL}) easily leads  from this formula.

\section{Evaluation of $x_{k_{i}k_{j}}$ and $x_{k_{i}k_{j}k_{l}}$}

From "string equation"
\begin{eqnarray}
Q(x,\{s\}) =0,
\end{eqnarray}
where
\begin{eqnarray}
Q(x,\{s\})=Q_{0}(x)+\sum\limits_{k=1}^{p-1} s_{k}Q_{k}(x)+...+
\sum\limits_{k_{i}=1}^{p-1}\frac{s_{k_{1}}...s_{k_{n}}}
{n!}Q_{k_{1}...k_{n}}(x)+...,
\end{eqnarray}
after differentiation,  we have
\begin{eqnarray}
\frac{\partial Q}{\partial x} dx +
\sum\limits_{k=1}^{p-1}\frac{\partial Q} {\partial s_{k}}ds_{k}+...+
\sum\limits_{k_{i}=1}^{p-1}\frac{\partial Q} {\partial
s_{k_{i}}}ds_{k_{i}}=0,
\end{eqnarray}
and we can see that
\begin{eqnarray}
Q' \frac{\partial x}{\partial s _{k_{i}}} +\frac{\partial
Q}{\partial s_{k_{i}}} =0, \quad \textrm{therefore} \quad x_{k_{i}}
= - \frac{Q_{k_{i}}}{Q'}.
\end{eqnarray}
In what follows we will use the  formula
\begin{eqnarray}
\frac{\partial Q_{k_{i}}}{\partial s_{k_{j}}} =
Q'_{k_{i}}x_{k_{j}}+Q_{k_{i}k_{j}}.
\end{eqnarray}
Evaluation of $x_{k_{i}k_{j}}$
\begin{multline}
x_{k_{i}k_{j}} = -\frac{\partial}{\partial s_{k_{j}}}
\left(\frac{Q_{k_{i}}}{Q'}\right) =
-\frac{Q'_{i}x_{k_{j}}+Q_{k_{i}k_{j}}}{Q'}+
\frac{Q_{k_{i}}(Q''x_{k_{j}}+
Q'_{k_{j}})}{(Q')^{2}}=\\=-\frac{Q_{k_{i}k_{j}}}{Q'}
+\frac{Q'_{k_{i}}Q_{k_{j}}+ Q'_{k_{j}}Q_{k_{i}}}{(Q')^{2}}-\frac{Q''
Q_{k_{i}} Q_{k_{j}}}{(Q')^{3}}.
\end{multline}
Evaluation of $x_{k_{i}k_{j}k_{l}}$
\begin{align}
&x_{k_{i}k_{j}k_{l}} = \frac{\partial}{\partial
s_{k_{l}}}\left(\frac{Q'_{k_{i}}Q_{k_{j}}+
Q'_{k_{j}}Q_{k_{i}}}{(Q')^{2}}-\frac{Q_{k_{i}k_{j}}} {Q'}
-\frac{Q''Q_{k_{i}}Q_{k_{j}}}{(Q')^{3}}\right) =\notag\\
&= \frac{(Q''_{k_{i}}x_{k_{l}}+Q'_{k_{i}k_{l}})
Q_{k_{j}}+Q'_{k_{i}}(Q'_{k_{j}} x_{k_{l}}+Q_{k_{j}k_{l}})+
(Q'_{k_{i}}x_{k_{l}}+Q_{k_{i}k_{l}})
Q'_{k_{j}}+Q_{k_{i}}(Q''_{k_{j}}x_{k_{l}}+
P'_{k_{j}k_{l}})}{(P')^{2}}
-\notag\\
&-\frac{2(Q'_{k_{i}}Q_{k_{j}}+Q_{k_{i}}
Q'_{k_{j}})(Q''x_{k_{l}}+Q'_{k_{l}})}
{(Q')^{3}}-\frac{Q'_{k_{i}k_{j}}x_{k_{l}}+
Q_{k_{i}k_{j}k_{l}}}{Q'}+\frac{Q_{k_{i}k_{j}}} {(Q')^{2}}
(Q''x_{k_{l}}+Q'_{k_{l}})-\notag\\
&-\frac{(Q_{k_{i}}Q_{k_{j}}
(Q'''x_{k_{l}}+Q''_{k_{l}})+(Q'_{k_{i}}x_{k_{l}} +Q_{k_{i}k_{l}})
Q''Q_{k_{j}}+Q_{k_{i}} Q''(Q'_{k_{j}}x_{k_{l}}+Q_{k_{j}k_{l}}))}
{(Q')^{3}}+\notag\\
&+\frac{3Q''Q_{k_{i}}Q_{k_{j}}} {(Q')^{4}}(Q''x_{k_{l}}+Q'_{k_{l}}).
\end{align}
Replace $x_{k_{i}}, x_{k_{j}}$ and $x_{k_{l}}$ in this formula we
obtain
\begin{align}
&x_{k_{i}k_{j}k_{l}} = -\frac{Q_{k_{i}k_{j}k_{l}}}{Q'}+
\frac{Q'_{k_{i}k_{j}}Q_{k_{l}}+
Q'_{k_{j}k_{l}}Q_{k_{i}}+Q'_{k_{l}k_{i}}Q_{k_{j}}} {(Q')^{2}}+
\frac{Q'_{k_{i}}Q_{k_{j}k_{l}}+Q'_{k_{j}}
Q_{k_{l}k_{i}}+Q'_{k_{l}}Q_{k_{i}k_{j}}}
{(Q')^{2}}-\notag\\
&-\frac{Q''_{k_{i}}Q_{k_{j}}Q_{k_{l}}
+Q''_{k_{j}}Q_{k_{l}}Q_{k_{i}}+
Q''_{k_{l}}Q_{k_{i}}Q_{k_{j}}}{(Q')^{3}}
-\frac{2(Q'_{k_{i}}Q'_{k_{j}}Q_{k_{l}}+Q'_{k_{j}}
Q'_{k_{l}}Q_{k_{i}}+
Q'_{k_{l}}Q'_{k_{i}}Q_{k_{j}})}{(Q')^{3}}-\notag\\
&-Q''\frac{Q_{k_{i}k_{j}} Q_{k_{l}}+Q_{k_{j}k}Q_{k_{i}}+
Q_{k_{l}k_{i}}Q_{k_{j}}}{(Q')^{3}}+
3Q''\frac{Q'_{k_{i}}Q_{k_{j}}Q_{k_{l}}
+Q'_{k_{j}}Q_{k_{l}}Q_{k_{i}}+
Q'_{k_{l}}Q_{k_{i}}Q_{k_{j}}}{(Q')^{4}}+\notag\\
&+Q''' \frac{Q_{k_{i}}Q_{k_{j}}Q_{k_{l}}}{(Q')^{4}}-3(Q'')^{2}
\frac{Q_{k_{i}}Q_{k_{j}}Q_{k_{l}}}{(Q')^{5}}.
\end{align}
Using  the notation of the  symmetrization we have
\begin{multline}
x_{k_{i}k_{j}k_{l}}=
-\frac{Q_{k_{i}k_{j}k_{l}}}{Q'}+\frac{Q'_{(k_{i}k_{j}}
Q_{k_{l})}}{(Q')^{2}} +\frac{Q'_{(k_{i}}Q_{k_{j}k_{l})}}{(Q')^{2}}-
\frac{Q''_{(k_{i}}Q_{k_{j}}Q_{k_{l})}}{(Q')^{3}}-
\frac{2Q'_{(k_{i}}Q'_{k_{j}}Q_{k_{l})}}{(Q')^{3}}
-Q''\frac{Q_{(k_{i}k_{j}}Q_{k_{l})}}{(Q')^{3}}+
\\+ 3Q'' \frac{Q'_{(k_{i}}Q_{k_{j}}Q_{k_{l})}}
{(Q')^{4}} +Q'''\frac{Q_{k_{i}}Q_{k_{j}}Q_{k_{l}}}
{(Q')^{4}}-3(Q'')^{2}\frac{Q_{k_{i}}Q_{k_{j}} Q_{k_{l}}}{(Q')^{5}}.
\end{multline}

\section{Evaluation of five-point numbers}
From equations (\ref{ZQ}) and (\ref{Corls}) one can calculate that
\begin{align}
&\mathcal{Z}_{k_{1}k_{2}k_{3}k_{4}k_{5}} = \frac{1}{2}\int_{-1}^{1}
\left(Q_{(k_{1}} Q_{k_{2}k_{3}k_{4}k_{5})}+Q_{(k_{1}k_{2}}
Q_{k_{3}k_{4}k_{5})}+
Q_{0}Q_{k_{1}k_{2}k_{3}k_{4}k_{5}}\right)dx+\notag\\
&+(Q_{(k_{1}k_{2}k_{3}}Q_{k_{4})}+Q_{(k_{1}k_{2}}Q_{k_{3})k_{4}}
+Q_{0}Q_{k_{1}k_{2}k_{3}k_{4}})x_{k_{5}}+\notag\\
&+(Q_{(k_{1}k_{2}k_{3}} Q_{k_{5})}+
Q_{(k_{1}k_{2}}Q_{k_{3})k_{5}}+Q_{0}
Q_{k_{1}k_{2}k_{3}k_{5}})x_{k_{4}}+\notag\\
&+(Q_{(k_{1}k_{2}k_{4}}Q_{k_{5})} +Q_{(k_{1}k_{2}} Q_{k_{4})k_{5}}+
Q_{0}Q_{k_{1}k_{2}k_{4}k_{5}})x_{k_{3}}+\notag\\
&+(Q'_{(k_{1}k_{2}}Q_{k_{3})}+Q_{(k_{1}k_{2}}Q'_{k_{3})}
+Q'_{0}Q_{k_{1}k_{2}k_{3}}
+Q_{0}Q'_{k_{1}k_{2}k_{3}})x_{k_{4}}x_{k_{5}}+\notag\\
&+(Q'_{(k_{1}k_{2}}Q_{k_{4})}+Q_{(k_{1}k_{2}}Q'_{k_{4})}+Q'_{0}Q_{k_{1}k_{2}k_{4}}
+Q_{0}Q'_{k_{1}k_{2}k_{4}})x_{k_{3}}x_{k_{5}}+\notag\\
&+(Q'_{(k_{1}k_{2}}Q_{k_{5})}+Q_{(k_{1}k_{2}}Q'_{k_{5})}
+Q'_{0}Q_{k_{1}k_{2}k_{5}}
+Q_{0}Q'_{k_{1}k_{2}k_{5}})x_{k_{3}}x_{k_{4}}+\notag\\
&+(Q_{(k_{1}k_{2}}Q_{k_{3})}+
Q_{0}Q_{k_{1}k_{2}k_{3}})x_{k_{4}k_{5}}+\notag\\
&+(Q_{(k_{1}k_{2}}Q_{k_{4})}+
Q_{0}Q_{k_{1}k_{2}k_{4}})x_{k_{3}k_{5}}+\notag\\
&+(Q_{(k_{1}k_{2}}Q_{k_{5})}+
Q_{0}Q_{k_{1}k_{2}k_{5}})x_{k_{3}k_{4}}+\notag\\
&+(Q''_{(k_{1}}Q_{k_{2})}+2Q'_{k_{2}} Q'_{k_{1}}+
Q''_{0}Q_{k_{1}k_{2}}+2Q'_{0}Q'_{k_{1}k_{2}}+
Q_{0}Q''_{k_{1}k_{2}})x_{k_{3}}x_{k_{4}}x_{k_{5}}+\notag\\
&+(Q'_{(k_{1}}Q_{k_{2})}+ Q'_{0}Q_{k_{1}k_{2}}+
Q_{0}Q'_{k_{1}k_{2}})(x_{(k_{3}k_{5}}x_{k_{4})})+(Q_{k_{1}}Q_{k_{2}}+Q_{0}
Q_{k_{1}k_{2}})x_{k_{3}k_{4}k_{5}}. \label{Zkkkkk}
\end{align}
All integrated terms are taken at the point $x=1$. Substituting in
this formula values for $x_{k_{i}}$, $x_{k_{i}k_{j}}$ and
$x_{k_{i}k_{j}k_{l}}$, calculated in Appendix B, after
simplification we find
\begin{align}
\mathcal{Z}_{k_{1}k_{2}k_{3}k_{4}k_{5}}  =& \frac{1}{2}\int_{-1}^{1}
\left(Q_{(k_{1}} Q_{k_{2}k_{3}k_{4}k_{5})}+
Q_{(k_{1}k_{2}}Q_{k_{3}k_{4}k_{5})}
+Q_{0}Q_{k_{1}k_{2}k_{3}k_{4}k_{5}}\right)dx-
\frac{Q_{(k_{1}k_{2}k_{3}}Q_{k_{4}}Q_{k_{5})}}
{Q_{0}'}-\notag\\
&-\frac{Q_{(k_{1}k_{2}}Q_{k_{3}k_{4}} Q_{k_{5})}}{Q_{0}'}+
\frac{Q'_{(k_{1}k_{2}}Q_{k_{3}}Q_{k_{4}} Q_{k_{5})}}{(Q_{0}')^{2}}
+\frac{Q'_{(k_{1}}Q_{k_{2}k_{3}} Q_{k_{4}}Q_{k_{5})}}{(Q'_{0})^{2}}-
\notag\\
&-\frac{Q''_{(k_{1}}Q_{k_{2}}Q_{k_{3}} Q_{k_{4}}Q_{k_{5})}}
{(Q'_{0})^{2}}- \frac{2Q'_{(k_{1}}Q'_{k_{2}}Q_{k_{3}}
Q_{k_{4}}Q_{k_{5})}} {(Q'_{0})^{3}}-
Q''_{0}\frac{Q_{(k_{1}k_{2}}Q_{k_{3}}Q_{k_{4}}
Q_{k_{5})}}{(Q'_{0})^{3}}+\notag\\
&+3Q''_{0} \frac{Q'_{(k_{1}}Q_{k_{2}}Q_{k_{3}}Q_{k_{4}}
Q_{k_{5})}}{(Q'_{0})^{4}}+\left(Q'''_{0}-\frac{3(Q''_{0})^{2}}{Q'_{0}}\right)
\frac{Q_{k_{1}}Q_{k_{2}} Q_{k_{3}}Q_{k_{4}}Q_{k_{5}}}{(Q'_{0})^{4}}.
\end{align}
Because  $Q_{k_{1}...k_{n}}(x) =\frac{d^{n-1}}{dx^{n-1}}L_{p-\sum k
-n}(x)$, using Appendix A, we find that
\begin{align}
&Q_{0}(1) = 0,\\
&Q'_{0}(1) = 1,\\
&Q_{k_{i}}(1) = 1,\\
&Q''_{0}(1) = \frac{p(p+1)}{2},\\
&Q'''_{0}(1) = \frac{1}{8}(p-1)p(p+1)(p+2),\\
&Q'_{k_{i}}(1)=F_{\Theta}(k_{i}-1),\\
&Q_{k_{i}k_{j}}(1)=F_{\Theta}(k_{ij}),\\
&Q_{k_{i}k_{j}k_{l}}(1)=H_{\Theta}(k_{ijl}),\\
&Q'_{k_{i}k_{j}}(1)=H_{\Theta}(k_{ij}-1),  \\
&Q''_{k_{i}}(1)=H_{\Theta}(k_{i}-2),
\end{align}
here new function were introduced
\begin{align}
&F_{\Theta}(k) = \frac{1}{2}(p-1-k)(p-2-k)\Theta(p-1-k),\\
&H_{\Theta}(k)=\frac{1}{2}F_{\Theta}(k)F_{\Theta}(k+2)=
\frac{1}{8}\Theta(p-1-k)\prod\limits_{r=1}^{4}(p-r-k).
\end{align}
After partial simplification we get
\begin{align}
\mathcal{Z}_{k_{1}k_{2}k_{3}k_{4}k_{5}} =& \frac{1}{2}\int_{-1}^{1}
\left(Q_{(k_{1}}Q_{k_{2}k_{3}k_{4}k_{5})}+
Q_{(k_{1}k_{2}}Q_{k_{3}k_{4}k_{5})}\right)dx+
\left(Q'''_{0}-3(Q''_{0})^{2}\right)- \notag\\
&+\sum\limits_{i=1}^{5}(3Q''_{0}Q'_{k_{i}}-Q''_{k_{i}})+
\sum\limits_{i<j}(Q'_{k_{i}k_{j}}-
2Q'_{k_{i}}Q'_{k_{j}}-Q''_{0}Q_{k_{i}k_{j}})+ \notag\\
&+
\sum\limits_{i,j,l}Q'_{k_{i}}Q_{k_{j}k_{l}}-\sum\limits_{i<j<l}Q_{k_{i}k_{j}k_{l}}-
\sum\limits_{i,j,k,l}Q_{k_{i}k_{j}}Q_{k_{l}k_{m}}.
\end{align}
Also introduce  the  functions
\begin{align}
&F(k)=\frac{1}{2}(p-1-k)(p-2-k),\\
&H(k)=\frac{1}{2}F(k)F(k+2)=
\frac{1}{8}\prod\limits_{r=1}^{4}(p-r-k),
\end{align}
which already doesn't depend from $\Theta(p-1-k)$. Divide  the
five-point correlation numbers on several parts
\begin{eqnarray}
\mathcal{Z}_{k_{1}k_{2}k_{3}k_{4}k_{5}}=
\mathcal{Z}_{k_{1}k_{2}k_{3}k_{4}k_{5}}^{(\textrm{I})}+
\mathcal{Z}_{k_{1}k_{2}k_{3}k_{4}k_{5}}^{(\textrm{J})}+
\mathcal{Z}_{k_{1}k_{2}k_{3}k_{4}k_{5}}^{(\textrm{1})}+
\mathcal{Z}_{k_{1}k_{2}k_{3}k_{4}k_{5}}^{(\textrm{2})},
\end{eqnarray}
where
\begin{align}
&\mathcal{Z}_{k_{1}k_{2}k_{3}k_{4}k_{5}}^{(\textrm{I})}
=\frac{1}{2}\int_{-1}^{1} Q_{(k_{1}}Q_{k_{2}k_{3}k_{4}k_{5})}dx,
&\mathcal{Z}_{k_{1}k_{2}k_{3}k_{4}k_{5}}^{(\textrm{J})}
=\frac{1}{2}\int_{-1}^{1} Q_{(k_{1}k_{2}}Q_{k_{3}k_{4}k_{5})}dx,
\end{align}
\begin{multline}
\mathcal{Z}_{k_{1}k_{2}k_{3}k_{4}k_{5}}^{(\textrm{1})}
=\sum\limits_{i=1}^{5}\left(\frac{3p(p+1)}{2}
F_{\Theta}(k_{i}-1)-H_{\Theta}(k_{i}-2)\right)-\\
-2\sum\limits_{i<j}F_{\Theta}(k_{i}-1) F_{\Theta}(k_{j}-1)
-\frac{p(p+1)(5p^{2}+5p+2)}{8},
\end{multline}
\begin{multline}
\mathcal{Z}_{k_{1}k_{2}k_{3}k_{4}k_{5}}^{(\textrm{2})}=
\sum\limits_{i<j}\left(H_{\Theta}(k_{ij}-1)
-F_{\Theta}(k_{ij})\frac{p(p+1)}{2}+F_{\Theta}(k_{ij})\sum\limits_{l\neq
i,j} F_{\Theta}(k_{l}-1)\right)-
\\-\sum\limits_{i<j<l}H_{\Theta}(k_{ijl})-\sum\limits_{i,j,l,m}
F_{\Theta}(k_{ij})F_{\Theta}(k_{lm}).
\end{multline}
Part of partition function
$\mathcal{Z}_{k_{1}k_{2}k_{3}k_{4}k_{5}}^{(\textrm{1})}$ has such a
difference, that one shouldn't take into account influence on it of
the factors $\Theta(a-b)$, due to  the  inequalities $k_{i}\leqslant
p-1$. So we can rewrite it by means of  the functions $F(k)$ and
$H(k)$, as follows
\begin{multline}
\mathcal{Z}_{k_{1}k_{2}k_{3}k_{4}k_{5}}^{(\textrm{1})}
=\sum\limits_{i=1}^{5}\left(\frac{3p(p+1)}{2}
F(k_{i}-1)-H(k_{i}-2)\right)-\\
-2\sum\limits_{i<j}F(k_{i}-1) F(k_{j}-1)
-\frac{p(p+1)(5p^{2}+5p+2)}{8}.
\end{multline}
Then using that
\begin{align}
&H_{\Theta}(k_{ijl}) = H(k_{ijl})(1- \Theta(k_{ijl}-p)), \notag\\
&H_{\Theta}(k_{ij}-1)= H(k_{ij}-1) (1-\Theta(k_{ij}-p)),\notag\\
&F_{\Theta}(k_{ij}) = F(k_{ij})(1- \Theta(k_{ij}-p)),\notag\\
&F_{\Theta}(k_{ij})F_{\Theta}(k_{lm})=
F(k_{ij})F(k_{lm})(1-(\Theta(k_{ij}-p)
+\Theta(k_{lm}-p)-\Theta(k_{ij}-p)\Theta(k_{lm}-p))),
\end{align}
let us write
$\mathcal{Z}_{k_{1}k_{2}k_{3}k_{4}k_{5}}^{(\textrm{2})}$, in the
following way
\begin{eqnarray}
\mathcal{Z}_{k_{1}k_{2}k_{3}k_{4}k_{5}}^{(\textrm{2})} =
\tilde{\mathcal{Z}}_{k_{1}k_{2}k_{3}k_{4}k_{5}}^{(\textrm{2})}+
\mathcal{Z}_{k_{1}k_{2}k_{3}k_{4}k_{5}}^{(\textrm{2}\Theta)},
\end{eqnarray}
where
\begin{multline}
\tilde{\mathcal{Z}}_{k_{1}k_{2}k_{3}k_{4}k_{5}}^{(\textrm{2})}=
\sum\limits_{i<j}\left(H(k_{ij}-1)
-F(k_{ij})\frac{p(p+1)}{2}+F(k_{ij})\sum\limits_{l\neq i,j}
F(k_{l}-1)\right)-
\\-\sum\limits_{i<j<l}H(k_{ijl})-\sum\limits_{i,j,l,m}
F(k_{ij})F(k_{lm}),
\end{multline}
which doesn't already depend on  the step-functions, and
\begin{multline}
\mathcal{Z}_{k_{1}k_{2}k_{3}k_{4}k_{5}}^{(\textrm{2}\Theta)} =\\
= \sum\limits_{m<n}\Theta(k_{mn}-p)
\left[F(k_{mn})\left(\sum\limits_{i<j}^{}F(k_{ij})
-\sum\limits_{l\neq m,n}F(k_{l}-1)+\frac{p(p+1)}{2}\right)
-H(k_{mn}-1)\right]+\\
+\sum\limits_{i<j<l}H(k_{ijl})\Theta(k_{ijl}-p)-
\sum\limits_{i,j,l,m}F(k_{ij})F(k_{lm})\Theta(k_{ij}-p)\Theta(k_{lm}-p),
\end{multline}
in which there is all dependence on $\Theta(a-b)$. Having
simplified, we have
\begin{multline}
\mathcal{Z}_{k_{1}k_{2}k_{3}k_{4}k_{5}}^{(\textrm{2}\Theta)} = -
\frac{1}{8}(2p-3-k)(2p-5-k)\sum\limits_{m<n}k_{ijl}^{mn}(k_{ijl}^{mn}+2)
\Theta(k_{mn}-p)+\\+\sum\limits_{i<j<l}
H(k_{ijl})[\Theta(k_{ijl}-p)+\Theta(k_{mn}-p)]
-\sum\limits_{i,j,l,m}F(k_{ij})F(k_{lm})\Theta(k_{ij}-p)\Theta(k_{lm}-p).
\end{multline}
On the other hand one can calculate
\begin{multline}
\mathcal{Z}_{k_{1}k_{2}k_{3}k_{4}k_{5}}^{(\textrm{1})}
+\tilde{\mathcal{Z}}_{k_{1}k_{2}k_{3}k_{4}k_{5}}^{(\textrm{2})}=\\
= \frac{1}{8}(4\sum_{i=1}^{5}k_{i}^{2}-k^{2}-2k -8)(2p-3-k)(2p-5-k)
- \sum\limits_{i<j<l}H(k_{ijl}).
\end{multline}
As  the  result we find that
\begin{multline}
\mathcal{Z}_{k_{1}k_{2}k_{3}k_{4}k_{5}}^{(\textrm{1})}
+\mathcal{Z}_{k_{1}k_{2}k_{3}k_{4}k_{5}}^{(\textrm{2})}=
\\
=\frac{1}{8}(4\sum_{i=1}^{5}k_{i}^{2}-k^{2}-2k -8-
\sum\limits_{m<n}k_{ijl}^{mn}(k_{ijl}^{mn}+2)
\Theta(k_{mn}-p))(2p-3-k)(2p-5-k)+\\
+\sum\limits_{i<j<l}H(k_{ijl})G_{1}-\sum\limits_{i,j,l,m}F(k_{ij})F(k_{lm})G_{2},
\label{Corup}
\end{multline}
where $G_{1}=\Theta(k_{ijl}-p)+ \Theta(k_{mn}-p)-1 \;$ and
$G_{2}=\Theta(k_{ij}-p)\Theta(k_{lm}-p)$.

In the even section we will need formulas
\begin{multline}
\frac{1}{2}\int_{-1}^{1}Q_{k_{i}k_{j}k_{l}k_{m}}(x)Q_{k_{n}}(x)dx =\\
=
\frac{1}{8}(k_{n}^{ijlm}-2)(k_{n}^{ijlm}-4)(2p-3-k)(2p-5-k)\Theta(k_{n}^{ijlm}-6),
\label{I1}
\end{multline}
\begin{multline}
\frac{1}{2}\int_{-1}^{1}Q_{k_{m}k_{n}}(x)Q_{k_{i}k_{j}k_{l}}(x)dx =\\
=\left(H(k_{ijl})-\frac{k^{mn}_{ijl}(k^{mn}_{ijl}+2)(2p-3-k)(2p-5-k)}
{8}\Theta(k_{mn}^{ijl}-2)\right)
\Theta(p-1-k_{ijl})\Theta(p-1-k_{mn}). \label{I2}
\end{multline}

\end{document}